\documentclass[runningheads,UKenglish]{llncs}
\usepackage[T1]{fontenc}
\usepackage[utf8]{inputenc}
\usepackage{babel}

\usepackage{graphicx}
\usepackage{wrapfig2}
\usepackage{amssymb}
\usepackage{mismath}
\usepackage{se2colors}
\usepackage[%
  csquotes=true,%
  booktabs=true,%
  siunitx=true,%
  minted=true,%
  selnolig=false,%
  widowcontrol=false,%
  microtype=true,%
  biblatex=false,%
  cleveref=true,%
]{se2packages}

\usepackage{paperdefs}

\usepackage{tikz}
\usetikzlibrary{%
  arrows,%
  arrows.meta,%
  automata,%
  backgrounds,%
  calc,%
  positioning,%
  shapes.geometric,%
  shapes.misc,%
  shapes.symbols,%
}
\usepackage{pgfplots}
\pgfplotsset{compat=newest, set layers=standard}

\usepackage{hyperref}

\newcommand{\dynamosadeONEPLUSONESUMdays}{1778.1402632309107}

\newcommand{\dynamosadeONEPLUSTENSUMdays}{1988.6846251225095}

\newcommand{\dynamosadeONEPLUSZEROSUMdays}{1547.5405450749818}

\newcommand{\dynamosadeTENPLUSONESUMdays}{1112.392357212868}

\newcommand{\dynamosadeZEROPLUSONESUMdays}{2353.160518240088}

\newcommand{\dynamosaGSdays}{4803.445761870926}
\newcommand{\TuningDynaMOSABaseCov}{88.90362233980623}
\newcommand{\TuningDynaMOSABaseRelCov}{71.32500087771653}
\newcommand{\TuningMIOBaseCov}{88.66517938289459}
\newcommand{\TuningMIOBaseRelCov}{77.44427001569858}

\newcommand{\miodeONEPLUSONESUMdays}{766.1656018823178}

\newcommand{\miodeONEPLUSTENSUMdays}{1151.9141142259982}

\newcommand{\miodeONEPLUSZEROSUMdays}{575.6587253258433}

\newcommand{\miodeTENPLUSONESUMdays}{384.7448329347043}

\newcommand{\miodeZEROPLUSONESUMdays}{384.04276336335744}

\newcommand{\mioGSdays}{9438.505603779457}

\begin{document}

\title{Hyperparameter Tuning for Search-based\\ Python Unit Test Generation}
\title{Search-based Hyperparameter Tuning for\\ Python Unit Test Generation}
\author{%
  Stephan Lukasczyk\inst{1,2}\orcidID{0000-0002-0092-3476} \and%
  Gordon Fraser\inst{2}\orcidID{0000-0002-4364-6595}%
}
\authorrunning{S. Lukasczyk and G. Fraser}
\institute{%
  JetBrains Research, Germany \and%
  University of Passau, Germany%
}

\maketitle

\begin{abstract}
  Search-based test-generation algorithms have countless configuration
  options. Users rarely adjust these options and usually stick to the
  default values, which may not lead to the best possible results.
  Tuning an algorithm's hyperparameters is a method to find better
  hyperparameter values, but it typically comes with a high demand of
  resources.
  Meta-heuristic search algorithms---that effectively solve the
  test-generation problem---have been proposed as a solution to also
  efficiently tune parameters.
  In this work we explore the use of differential evolution as a means
  for tuning the hyperparameters of the DynaMOSA and MIO
  many-objective search algorithms as implemented in the \pynguin
  framework.
  Our results show that significant improvement of the resulting test
  suite's coverage is possible with the tuned DynaMOSA algorithm and
  that differential evolution is more efficient than basic grid
  search.

\end{abstract}

\section{Introduction}\label{sec:intro}

Many algorithms allow adjusting their behaviour by providing
parameters, so-called \emph{hyperparameters}, to the user.  The more
complex an algorithm is or the more hyperparameters it exposes, the
harder it is to find values that imply optimal behaviour of the
algorithm.  Tuning of the hyperparameters is one way to find optimal
values, but it is usually time-consuming~\cite{MRN14}.  Studies show
that about \perc{80} of the literature in Software Engineering, Data
Mining, and Defect Prediction—fields that heavily use complex
algorithms—do not mention tuning at all, although an improvement of up
to \perc{60} is possible~\cite{FMS16}.  In many cases the cost of
tuning does not seem to be worth the effort~\cite{VQ21}.

One category of such complex and heavily configurable algorithms are
\emph{evolutionary algorithms}~(EAs).  They are popular for many
optimisation problems, \eg, search-based test generation~\cite{McM04}.
Various algorithms, such as DynaMOSA~\cite{PKT18b} or
MIO~\cite{Arc18}, implemented in mature tools, such as
\evosuite~\cite{FA11} for Java or \pynguin~\cite{LF22} for Python, are
representative for this technique.  EAs consist of various operators,
each with numerous hyperparameters with significant
influence~\cite{HHL+09}.  While tools typically come with default
hyperparameter values that allow their off-the-shelf usage, it is an
open question whether these default values are actually suitable for a
problem at hand.

Among many existing tuning approaches and techniques~\cite{HLY20}, the
\emph{differential evolution}~\cite{SP97} search algorithm has been
extensively studied and is reported to provide very good tuning
results in short time~\cite{FMS16,FNM16,FM17}.
In this paper we therefore explore how well differential evolution can
tune a set of hyperparameters of the DynaMOSA and MIO algorithms as
implemented in the \pynguin test-generation framework for Python.  To
assess the improvements and computational costs, we compare against a
baseline of \emph{grid search}~\cite{BB12}.
In detail, the contributions of this paper are the following:
\begin{enumerate}
\item We conduct a large scale hyperparameter tuning experiment for
  the DynaMOSA and MIO algorithms in the \pynguin test-generation
  framework.
  \item We tune each algorithm's hyperparameters with two tuning
    algorithms, differential evolution and grid search.
\end{enumerate}

The results confirm that hyperparameter tuning is a time-consuming
task, with our experiments consuming almost \(72~\mathrm{years}\) of
runtime.  For both tuned test-generation algorithms the tuning with
differential evolution was faster than with grid search.  However,
only for the DynaMOSA algorithm our tuning lead to a significantly
better performance in terms of coverage of the resulting test suites.
While this improvement can justify the effort, this depends on the use
case: The tuning of MIO confirms that default values can still lead to
reasonable results~\cite{AF13}.

\section{Background}\label{sec:background}

\subsection{Grid Search}\label{sec:background-grid}

Exhaustive tuning strategies explicitly and systematically check all
possible combinations of hyperparameter values, whether a combination
satisfies the desired properties.  This approach is also known as
\emph{brute force}.  A standard technique to exhaustively explore the
space of combinations is \emph{grid search}~\cite{BB12}, which
considers every possible combination of a given set of
hyperparameters~\cite{BB12}.  The number of joint values in the grid
grows exponentially with the number of hyperparameters to tune
(\emph{curse of dimensionality}~\cite{Bel57,Bel61}).  While this can
cause enormous runtime requirements, the algorithm is inherently
parallel~\cite{HS12} because all individual tuning runs are
independent of each other.
The computation of the grid requires discrete values for each
hyperparameter.  For hyperparameters with real values this requires
the user to decide on a discretisation, which may cause that one
misses the optimal value for a hyperparameter.

\subsection{Differential Evolution}\label{sec:background-de}

Differential evolution attempts to find the global optimum of non-linear,
non-convex, multi-modal, and non-differentiable functions defined in the
continuous parameter space~\cite{SP97}.  Its structure is similar to an EA,
however, it generates an offspring~\(\vec{x}_0'\)~(an \(n\)-dimensional vector
in the parameter space) by randomly choosing three distinct
individuals~\(\vec{x}_r\), \(\vec{x}_s\), and \(\vec{x}_t\) from the population
and combining them with a scale factor~\(F\), such that
\(
  \vec{x}_0' = \vec{x}_t + F(\vec{x}_r - \vec{x}_s)
\)~\cite{NT10}.
Given an objective function~\(f\), differential evolution aims to find the
global minimum of~\(f(\vec{x})\) in the decision space~\cite{NT10}.
The simplicity of differential evolution and its robustness and versatility
allow engineers, practitioners, and researchers to apply and adapt it in
countless ways.  Various literature surveys, \eg~\cite{NT10,DMS16},
present the large flexibility of the algorithm and its applications.

\subsection{DynaMOSA and MIO Algorithms}\label{sec:background-algorithms}

DynaMOSA~\cite{PKT18b} is a state-of-the-art EA that aims to optimise many
objectives at the same time.  In the context of test generation, with branch
coverage as an optimisation goal, each branch is an objective to the algorithm.
By incorporating the hierarchical structure of the program under test, DynaMOSA
is able to optimise for only those goals that are reachable at a given time.
The algorithm uses an archive as a second population, which stores those
individuals that successfully covered an optimisation goal.

MIO~\cite{Arc18} explicitly targets subjects with thousands of
(independent) optimisation goals.  It combines the simplicity of a
\((1+1)\)EA with feedback-directed target selection, a dynamic
population, a dynamic exploration/exploitation switch, and archives to
store populations for each goal.  The idea of the phase switch is that
in the beginning exploration helps discovering large parts of the
search space, whereas later in the process, exploitation allows to
focus on better results.

\subsection{\pynguin}\label{sec:background-pynguin}

\Pynguin~\cite{LF22} is a state-of-the-art unit test generation tool for the
Python programming language.  It implements various standard test-generation
algorithms, such as, Whole Suite~\cite{FA13}, DynaMOSA~\cite{PKT18b}, or
MIO~\cite{Arc18}.  \Pynguin aims to generate Python unit tests that reach
high branch coverage for given subject systems.

\section{Search-based Hyperparameter Tuning for SBST}\label{sec:approach}

The DynaMOSA~\cite{PKT18b} implementation in \pynguin showed the best
performance in previous work~\cite{LKF23} compared to the other
test-generation algorithms implemented in the framework, including
\pynguin's implementation of MIO~\cite{Arc18}.  The implementation of
these algorithms in \pynguin uses parameter values taken from
\evosuite~\cite{FA11}, a state-of-the-art unit-test generation tool
for Java.

\subsection{General Hyperparameters}\label{sec:approach-general-parmeters}

While both DynaMOSA and MIO are EAs, they are considerably different
in how they are built and what operators they use.  Both algorithms
share that they use \emph{test cases} as their chromosomes, in the
form of sequences of statements.  Because shorter test cases are
considered more readable and understandable for
developers~\cite{DCF+15}, tuning the maximum
\emph{chromosome length}, \ie, the number of statements in a chromosome, is a
natural choice.  Both algorithms handle their chromosome population differently,
thus, we decided to use different size ranges for them.  Additionally, while
differential evolution works on continuous intervals of floating-point numbers,
grid search requires a discretisation of the values.  We use the
differential-evolution implementation from the Python library \pypi{scipy} for
this study, which also supports intervals over integers.
\begin{itemize}
  \item \emph{DynaMOSA:} We constrain the chromosome length to the
    interval~\(\interval{5}{100}\subset\mathbb{N}\) for differential evolution;
    for grid search, we use~\(\{5, 10, 25, 50, 100\}\).
  \item \emph{MIO:} We bind the chromosome length to the
    interval~\(\interval{10}{50}\subset\mathbb{N}\) for differential evolution;
    we use~\(\{10, 25, 50\}\) for grid search.
\end{itemize}

Mutation is one of the three standard operators of an EA\@.  Usually,
it is only applied once in every iteration of the algorithm and due to
its stochastic nature there is only a certain probability for a
change.  MIO, however, allows applying mutation more than once to an
individual before sampling a new one.  We transfer the idea of
applying mutation more than once to DynaMOSA, thus allowing to tune
the \emph{number of mutations} with the following value ranges and
discretisations:
\begin{itemize}
  \item \emph{DynaMOSA:} We constrain the number of mutations to the
    interval~\(\interval{0}{25}\subset\mathbb{N}\) for differential evolution;
    for grid search, we use~\(\{0, 1, 5, 10, 25\}\).
  \item \emph{MIO:} Since MIO does not use crossover at all, we do not want to
    disable mutation.  Thus, we constrain the number of mutations for
    differential evolution to~\(\interval{1}{25}\subset\mathbb{N}\); for grid
    search, we use~\(\{1, 10, 25\}\).  Note that these intervals apply to both
    phases of MIO\@; we tune the number of mutations for both independently.
\end{itemize}

Since we cannot assume that an EA will find an optimal solution, it is
necessary to define a \emph{search budget}.  This often is the only
parameter a user of a tool will adjust, because it is directly
understandable to them~\cite{AF13}.  While different budget types,
\eg, algorithm iterations, are possible, we decided to use a timeout
of \qty{180}{\second} because a user usually wants to control the
execution time of a tool.

\subsection{DynaMOSA-specific Parameters}\label{sec:approach-dynamosa-params}

A DynaMOSA-specific parameter is the \emph{population size}, i.e., the
number of individuals in the EA's population.  A large size allows for
more diversity in the population, which can escape local optima in the
fitness landscape easier.  However, a large size can also slow down
convergence towards the global optimum~\cite{AF13}.  For differential
evolution, we constrain the population size
to~\(\interval{4}{200}\subset\mathbb{N}\), for grid search
to~\(\{4, 10, 50, 100, 200\}\), same as in previous work~\cite{AF13}.

The \emph{crossover rate} specifies the probability that two selected
individuals are crossed over.  For grid search, we
use~\(\{0, 0.25, 0.5, 0.75, 1\}\), for differential
evolution~\(\interval{0}{1}\).  For the selection of individuals, we choose
between rank and tournament selection; for grid search, we set the rank bias
to~\(\{1.2,1.7\}\) and for differential evolution to~\(\interval{1.01}{1.99}\).
The tournament size is bound to~\(\{2,7\}\) for grid search
and~\(\interval{1}{20}\subset\mathbb{N}\) for differential evolution.

\subsection{MIO-specific Parameters}\label{sec:approach-mio-params}

Characteristic to MIO is its switch between exploration and exploitation phases.
The former shall explore large parts of the fitness landscape whereas the latter
shall fine-tune the individuals in the population.  We allow degenerating MIO
to only use either exploration or exploitation by using~\(\interval{0}{1}\) for
differential evolution and~\(\{0, 0.25, 0.5, 0.75, 1\}\) for grid search.

\emph{Number of tests per target} is similar to the population size.
For each optimisation goal, MIO holds one population in its archive.
During exploration we allow numbers
from~\(\interval{1}{25}\subset\mathbb{N}\) for differential evolution
and~\(\{1, 10, 25\}\) for grid search, respectively.  MIO only keeps
one test per target in the exploitation phase~\cite{Arc18}.

Similarly, during exploration MIO can either select a test from the archive or
sample a new one.  The probability of this sampling is set
to~\(\interval{0}{1}\) for differential evolution
and~\(\{0, \sfrac{1}{3}, \sfrac{2}{3}, 1\}\) for grid search.  In the
exploitation phase, the probability is always~\(0\) because the algorithm shall
only refine existing test cases.

\subsection{Fitness Function for the Tuner}\label{sec:approach-tuner-fitness}

Since \pynguin aims to generate test suites with high coverage, the
coverage of the final test suite provides an obvious choice for
selecting the best set of hyperparameters. However, it may also be
desirable to achieve coverage as quickly as possible, which might not
be reflected by the final coverage. The area under the curve of the
coverage achieved over time therefore provides an alternative metric.
This lets us define five different fitness functions for the tuning
algorithms: solely coverage~(denoted as \DEconfig{1}{0}), only area
under curve~(\DEconfig{0}{1}), the sum of coverage and area under
curve~(\DEconfig{1}{1}), and two weighted sums: ten times coverage
plus area under curve~(\DEconfig{10}{1}) and vice
versa~(\DEconfig{1}{10}).  Note that the aforementioned notation
refers to the respective objective function used for differential
evolution.
Grid search also requires a metric to choose the best configuration after all
configurations from the grid have been explored successfully.  We decided to use
the same metrics as for differential evolution's objective function.  We use
\GSconfig{\(\alpha\)}{\(\beta\)} to denote these functions, respectively.

\section{Empirical Evaluation}\label{sec:evaluation}

To evaluate differential evolution as a tuning algorithm for DynaMOSA
and MIO in \pynguin with grid search as a baseline tuning algorithm,
we investigate the following research questions:
\begin{itemize}
\item \textbf{RQ1:} How much do the tuning algorithms improve the performance?
\item \textbf{RQ2:} How do the algorithms compare in terms of computational costs?
\end{itemize}

\subsection{Experimental Setup}\label{sec:eval-setup}

We conducted an empirical evaluation using a set of modules as
evaluation subjects from previous work on \pynguin~\cite{LKF23}.  In
line with previous work on tuning~\cite{AF13} and recent work on
search-based unit test generation (e.g.,~\cite{LPL+23}), we
incorporate only those modules into our evaluation subjects where
\pynguin uses up the entire search budget and achieves between
\perc{80} and \perc{100} branch coverage, since the choice of
parameter values will have no influence on code that is either trivial
or impossible to cover for \pynguin. This leads to \exnum{101}
modules, which we split into a training and test set using an
\perc{80} split.

We use \pynguin in git revision \texttt{fd9a6e96} for the evaluation and
execute it inside Docker containers for process isolation.  We limit the search
budget of \pynguin to \qty{180}{\second} and disable its assertion generation
and test-code export.  To minimise the influence of randomness, we execute
\pynguin on each
subject \exnum{15} times.  Both values are in line with previous
research~\cite{AF13}.  All runs were executed on dedicated compute servers
equipped with an AMD EPYC 7443P CPU and \qty{256}{\giga\byte} RAM\@.  We assign
one CPU core and \qty[round-mode=none]{4}{\giga\byte} RAM to each execution.
All measured values are rounded to three significant digits.  We use the
Vargha-Delaney effect size \(\hat{A}_{12}\)~\cite{VD00} and the Mann-Whitney-U
test~\cite{MW47} at \(\alpha=0.05\) to compare two result sets, as recommended
by the literature~\cite{AB14}.

\subsection{Threats to Validity}\label{sec:eval-threats}

Threats to its \emph{internal validity} may result from bugs in
\Pynguin or our experiment framework, although both have been tested
carefully.  The stochastic nature of the test-generation algorithms
and differential evolution can cause results by chance.  We repeat
each \pynguin experiment \exnum{15} times with different random seeds
and follow rigorous statistical procedures to evaluate the results.
However, we do not repeat executions of differential execution due to
the already huge resource effort.
The study's \emph{external validity} is influenced by the small size
of \exnum{101} modules for the study, which may not allow to
generalise the results.  In general, the results from tuning are
inherently tied to the used subjects.  Another set of subjects might
result in different optimal configurations.
The fact that our evaluation only relies on branch coverage implies a threat to
\emph{construct validity}: practitioners might be interested in other metrics,
too, such as test-suite size or readability of test cases.  We consider these
factors negligible for this study but admit that one might need to consider them
in other use cases.

\subsection{RQ1: Performance Improvement due to Tuning}

\begin{figure}[t]
  \centering
  \includegraphics[width=0.95\textwidth]{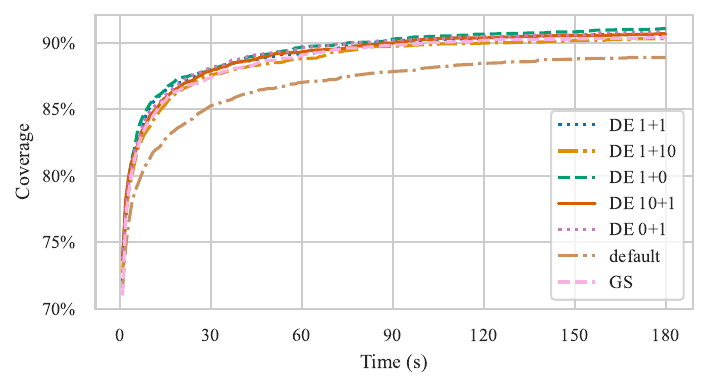}
  \caption{\label{fig:tuning-dynamosa-coverage-over-time}%
    Development of the mean branch coverage over time for the best DynaMOSA
    configurations.
  }
\end{figure}

\begin{figure}[t]
  \centering
  \includegraphics[width=0.95\textwidth]{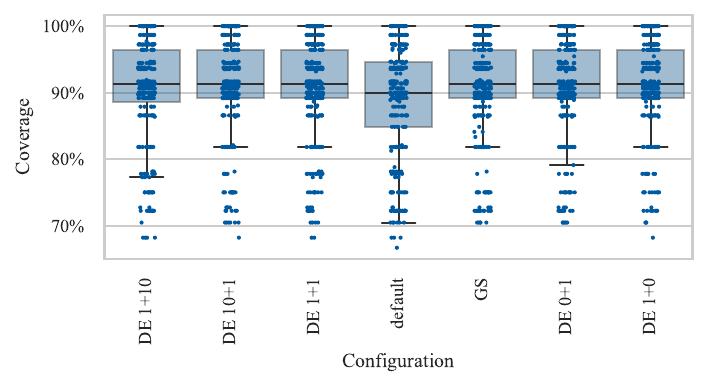}
  \caption{\label{fig:tuning-dynamosa-coverage-distributions}%
    Coverage distributions for the best DynaMOSA configurations.
  }
\end{figure}

\begin{table}[t]
  \centering
  \tabcaption{\label{tab:tuning-dynamosa-settings}%
    Best hyperparameter settings for DynaMOSA per tuning fitness function.
  }
  \begin{tabular}{ l S[round-mode=none] S S[round-mode=none] S[round-mode=none] S l S[round-mode=none] } \toprule
{Configuration}
  & \( l_c \)
    & \( P_c \)
      & \( n_m \)
        & \( N \)
          & \( \mathfrak{B} \)
            & {Selection}
              & \( \mathfrak{S} \)\\
\midrule
DE 1+1
  & 53
    & 0.7371902799042689
      & 3
        & 18
          & 1.3880136909602547
            & rank
              & 4 \\
DE 1+10
  & 39
    & 0.6761984293507988
      & 2
        & 18
          & 1.340019211181281
            & rank
              & 12 \\
DE 1+0
  & 48
    & 0.6480085675338735
      & 3
        & 10
          & 1.681839842637804
            & rank
              & 4 \\
DE 10+1
  & 45
    & 0.5725919133789719
      & 4
        & 8
          & 1.4399447184211887
            & rank
              & 12 \\
DE 0+1
  & 46
    & 0.5485404465990583
      & 3
        & 10
          & 1.3417410413300472
            & rank
              & 3 \\
default
  & 40
    & 0.75
      & 1
        & 50
          & 1.7
            & tournament
              & 5 \\
GS
  & 100
    & 0.75
      & 1
        & 4
          & 1.2
            & rank
              & 5 \\
\bottomrule
\multicolumn{8}{l}{\footnotesize chromosome length~\(l_c\), crossover rate~\(P_c\), number of mutations~\(n_m\),}\\
\multicolumn{8}{l}{\footnotesize population size~\(N\), rank bias~\(\mathfrak{B}\), tournament size~\(\mathfrak{S}\)}
\end{tabular}

\end{table}

\Cref{fig:tuning-dynamosa-coverage-over-time} shows the development of
the coverage over \emph{DynaMOSA}'s generation time of
\qty{180}{\second} for the configurations produced by the different
tuning algorithms and objectives.  All differential evolution
configurations achieve similar coverage values in the end; for grid
search, applying the different tuning fitness functions always yielded
the same configuration, which we denote as \emph{GS}.  However, all
configurations show a higher final coverage than with the algorithm's
default settings.  The similarities between the tuned algorithm
variants and the difference to the default can also be seen from the
coverage
distributions~(\cref{fig:tuning-dynamosa-coverage-distributions}).
While all tuned algorithm variants perform similar, \DEconfig{1}{0} is
the best in this experiment.

\Cref{tab:tuning-dynamosa-settings} shows the best hyperparameter
settings for each tuning method for DynaMOSA\@.  All configurations use
rank selection, in contrast to the default configuration.  Note that
while the tournament size~(column \(\mathfrak{S}\)) changes due to how
differential evolution applies its changes, it does not affect the
result because it is never used by rank selection.
For all configurations, except \DEconfig{1}{10}, a larger chromosome
length is preferable, whereas the crossover probability can be smaller
and more mutations are beneficial.  Furthermore, a smaller rank bias
is also preferable as is a smaller population size.

\begin{figure}[t]
  \centering
  \includegraphics[width=0.95\textwidth]{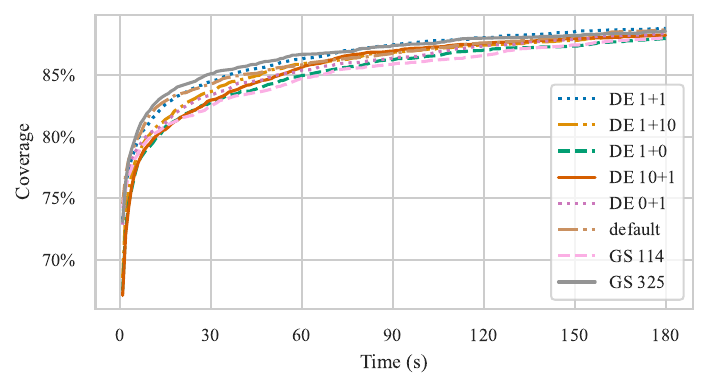}
  \caption{\label{fig:tuning-mio-coverage-over-time}%
    Development of the mean branch coverage over time for the best MIO
    configurations.
  }
\end{figure}

\begin{figure}[t]
  \centering
  \includegraphics[width=0.95\textwidth]{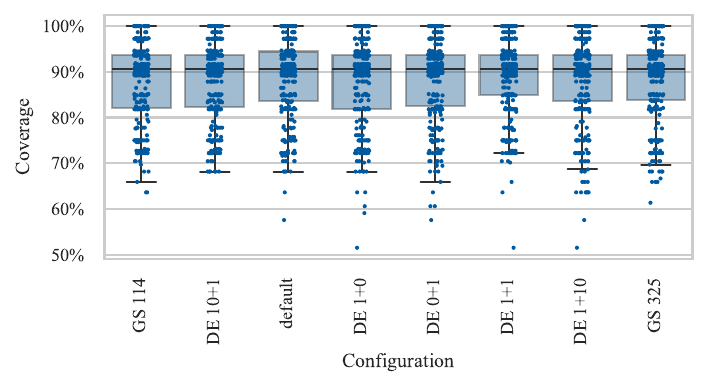}
  \caption{\label{fig:tuning-mio-coverage-distributions}%
    Coverage distributions for the best MIO configurations.
  }
\end{figure}

For \emph{MIO}, we see different results than for DynaMOSA\@.  While
the coverage over time~(\cref{fig:tuning-mio-coverage-over-time})
evolves differently for the different configurations in the first
\qty[round-mode=none]{30}{\second} to \qty[round-mode=none]{60}{\second},
one can barely distinguish them after
\qty{180}{\second}.  We note that there are now two grid-search
configurations: \mbox{GS\,114}, which is best for \GSconfig{1}{0} and
\GSconfig{10}{1}, and \mbox{GS\,325}, which is best for the other
grid-search configurations.  The coverage distributions only differ
slightly~(\cref{fig:tuning-mio-coverage-distributions}).
\Cref{tab:tuning-mio-settings} shows the corresponding hyperparameter
settings.  Note that the number of tests per target~\(\abs{T_k}\) and
the probability to randomly sample a new test case~\(P_r\) are fixed
to~\(\abs{T_k} = 1\) and \(P_r = 0\) as a characteristic property of
the exploitation~\cite{Arc18}.
Outstanding is \mbox{GS\,114}, which has no exploration phase at all; other
configurations also have the phase switch after less than \perc{10}; they all
favour exploitation over exploration.

\begin{table}[t]
  \centering
  \tabcaption{\label{tab:tuning-mio-settings}%
    Best hyperparameter settings for MIO per tuning fitness function.
  }
  \begin{tabular}{ l S[round-mode=none] S S[round-mode=none] S S[round-mode=none] S[round-mode=none] S S[round-mode=none] } \toprule
{Configuration}
  & \( l_c \)
    & \( F \)
      & \multicolumn{3}{c}{Exploration}
            & \multicolumn{3}{c}{Exploitation}\\
  & & & \( \abs{T_k} \)
        & \( P_r \)
          & \( n_m \)
            & \( \abs{T_k} \)
              & \( P_r \)
                & \( n_m \)\\
\midrule
DE 1+1
  & 48
    & 0.761911379260028
      & 24
        & 0.6725632331222376
          & 1
            & 1
              & 0
                & 5 \\
DE 1+10
  & 12
    & 0.09257220676705963
      & 16
        & 0.7281693389867376
          & 6
            & 1
              & 0
                & 1 \\
DE 1+0
  & 28
    & 0.09293011967714027
      & 1
        & 0.02319032555223932
          & 6
            & 1
              & 0
                & 2 \\
DE 10+1
  & 17
    & 0.2952122304290511
      & 24
        & 0.31171486653710284
          & 7
            & 1
              & 0
                & 4 \\
DE 0+1
  & 35
    & 0.07844694968476967
      & 16
        & 0.3325342365940621
          & 1
            & 1
              & 0
                & 13 \\
default
  & 40
    & 0.5
      & 10
        & 0.5
          & 1
            & 1
              & 0
                & 10 \\
GS 114
  & 10
    & 0.0
      & 10
        & 0.666666666666666
          & 1
            & 1
              & 0
                & 10 \\
GS 325
  & 10
    & 0.25
      & 1
        & 0.333333333333333
          & 1
            & 1
              & 0
                & 1 \\
\bottomrule
\multicolumn{9}{l}{\footnotesize chromosome length~\(l_c\), phase switch~\(F\), number of tests per target~\(\abs{T_k}\), pro-}\\
\multicolumn{9}{l}{\footnotesize bability of random sampling a new test case~\(P_r\), number of mutations~\(n_m\)}
\end{tabular}

\end{table}

\rqsummary{RQ1}{%
  All tuned DynaMOSA configurations improve over the default
  configuration, but there is no improvement for MIO\@. Differential
  evolution achieves better results than grid search.  }

\emph{Discussion:}
All tuned variants of DynaMOSA show an improvement in coverage over the default
configuration, which indicates that the default hyperparameter values
are not optimal. %
\Cref{tab:tuning-dynamosa-coverage-effect-sizes} reports the mean branch and
relative~\cite{AF13} coverages for the tuned configurations and compares them to
the default settings.  All tuned configurations yield significantly higher mean
coverage.

\begin{table}[t]
  \centering
  \tabcaption{\label{tab:tuning-dynamosa-coverage-effect-sizes}%
    Branch and relative coverage for DynaMOSA compared to the default configuration, together
    with their respective effect sizes and \pvalue{}s.
  }
  \begin{tabular}{ l S S S S S S } \toprule
{Configuration}
  & \multicolumn{3}{c}{Branch Coverage}
        & \multicolumn{3}{c}{Relative Coverage}\\
  & {mean \%}
    & {\(\hat{A}_{12}\)}
      & {\pvalue}
        & {mean \%}
          & {\(\hat{A}_{12}\)}
            & {\pvalue}\\ \midrule
default & \TuningDynaMOSABaseCov&—&—&\TuningDynaMOSABaseRelCov&—&—\\
DE\,1+1 & 90.63563199476408 & 0.5569280234427464 & 0.013499972073487358 & 83.6867431931126 & 0.5779237296758307 & 4.775510296430763e-05 \\
DE\,1+10 & 90.31365450978122 & 0.5481768852375881 & 0.036412513848072044 & 82.12744927030641 & 0.5699274811095897 & 0.0002742436238562653 \\
DE\,1+0 & 91.09883707001708 & 0.5674273543283128 & 0.0034029230052969313 & 85.75213675213675 & 0.5908007505451595 & 1.539092433498424e-06 \\
DE\,10+1 & 90.6887262050377 & 0.5572848521730311 & 0.01283608533747185 & 82.28257456828885 & 0.5712967189005528 & 0.00021321592625228258 \\
DE\,0+1 & 90.83519481987874 & 0.5640853998681474 & 0.005381411415051642 & 84.70695970695971 & 0.5833561539631827 & 1.2511942362198559e-05 \\
GS & 90.41635306486071 & 0.5464881586287337 & 0.04351131188755646 & 81.89790104075819 & 0.5620670419392464 & 0.0014403060928676928 \\
\bottomrule
\end{tabular}

\end{table}

Comparing the DynaMOSA configurations reveals that a larger maximum
chromosome length and a significantly smaller population size are
beneficial.  Usually, one would expect a larger population size, which
allows for higher diversity in the population.  Still, for grid
search, a population of only four individuals is sufficient to
outperform the default configuration with \exnum{50} individuals.
However, in this case, the diversity might come from the larger
chromosome length: the test cases have not been minimised after
generation, thus, it is possible that individual test cases actually
cover more than one focal method.  Additionally, other work that
applied tuning to the population-size parameter of DynaMOSA reports
that a smaller-than-default size is better~\cite{CGA18}.
Furthermore, introducing the possibility to run more than one mutation per
evolution step seems to be beneficial.  This, together with the smaller
crossover rates indicates that mutation is actually more important for
DynaMOSA's performance than one might expect.

None of the tuned variants of MIO show an improvement over the default
configuration.  Worse, if one compares the \(\hat{A}_{12}\) effect sizes of the
variants and the default configuration, this effect is slightly in favour of
the default when measuring it on mean branch coverage; for relative coverage,
two configurations have a slightly positive, although not significant, effect
towards the tuned configurations.

\begin{table}[t]
  \centering
  \tabcaption{\label{tab:tuning-mio-coverage-effect-sizes}%
    Branch and relative coverage for MIO compared to the default configuration,
    together with their respective effect sizes and \pvalue{}s.
  }
  \begin{tabular}{ l S S S S S S } \toprule
{Configuration}
  & \multicolumn{3}{c}{Branch Coverage}
        & \multicolumn{3}{c}{Relative Coverage}\\
  & {mean \%}
    & {\(\hat{A}_{12}\)}
      & {\pvalue}
        & {mean \%}
          & {\(\hat{A}_{12}\)}
            & {\pvalue}\\ \midrule
default & \TuningMIOBaseCov&—&—&\TuningMIOBaseRelCov&—&—\\
DE\,1+1 & 88.80259197155094 & 0.4977122700932225 & 0.9209023428697198 & 78.29565672422815 & 0.5087528344671202 & 0.6758766994009042 \\
DE\,1+10 & 88.25593670719844 & 0.49386243386243384 & 0.7896717050817397 & 76.27426013140298 & 0.5020005039052658 & 0.9239402186785315 \\
DE\,1+0 & 87.98315119846767 & 0.4867220962459058 & 0.563725404238661 & 72.74358974358974 & 0.47514235323759135 & 0.23915394161934966 \\
DE\,10+1 & 88.24449950734619 & 0.48621315192743764 & 0.5488884911298479 & 74.6648642362928 & 0.48309901738473165 & 0.4236451781372593 \\
DE\,0+1 & 88.35431006093782 & 0.496381960191484 & 0.8751179362257703 & 75.26612012326297 & 0.48853615520282184 & 0.5855873805856653 \\
GS\,114 & 88.05300613769005 & 0.48086671705719325 & 0.4054719909470865 & 73.99395313681026 & 0.47733938019652306 & 0.28487920461634264 \\
GS\,325 & 88.56541382847371 & 0.4981607457797934 & 0.936410823712083 & 76.92703064131634 & 0.4966137566137566 & 0.8723864835046813 \\
\bottomrule
\end{tabular}

\end{table}

No tuned configuration shows improvement over the default for MIO\@.
The choice of hyperparameters and their value ranges might limit the
results; selecting them differently could change the result.  However,
the chosen hyperparameters are those that are directly related to the
MIO algorithm, and that are also similar to the hyperparameters chosen
for the DynaMOSA tuning, to also allow for some inter-algorithm
reasoning.  The wide range of values for the different
hyperparameters, see \cref{tab:tuning-mio-settings}, however, does not
directly allow the conclusion that the selection of the value ranges
imposes this result.  The MIO algorithm performs very similar,
independently of its settings.

The results of both algorithms, DynaMOSA performing better than MIO, are in line
with previous research~\cite{CGA18,LKF23}.  However, there also exists research
that states the opposite~\cite{Arc18}.  We hypothesise that the choice of
subjects plays a large role in this: the subjects we chose are from libraries.
Since the test generation works on module level the number of goals per
individual subject is not huge.  MIO was designed for subjects with many
\emph{independent} goals, \eg, to generate system tests.  Its computationally
cheap evolution step allows it to explore much larger parts of the fitness
landscape in the same time than the computationally costly DynaMOSA\@.
Important to this hypothesis is probably the term \emph{independent}: DynaMOSA's
optimisation over its predecessor, MOSA~\cite{PKT15}, was that it used a
structural property of the subject, the nesting of branches: nested branches are
only relevant as goals for the search if the surrounding branch was covered,
\ie, there exists a test case in the population that evaluates the branch's
condition to true.  MIO's underlying assumption is the independence of the
goals, which is obviously not given for nested branches.  MIO always takes all
goals into account, although the nested ones might not even be reachable and
thus irrelevant.  This can cause a considerable waste of computational
resources. However, future research is necessary to test this hypothesis.

\subsection{RQ2: Tuning Costs}

Hyperparameter tuning is a time-consuming task~\cite{EHM99}.  Thus, if one
wants to invest the resources, it is advisable to choose a tuning method that
ekes the resources and yields high-quality results.  We study the amount of
resources consumed in order to provide recommendations for the best tuning
method.  We report \pynguin's total runtime here, which does not include, \eg,
the overhead for running Docker.  While we admit that the overhead is
considerable, we assume it as almost constant and not influencing the tuning
itself.

For the \emph{MIO} tuning, grid search consumed a total amount of
\qty{\mioGSdays}{\day}.  Because of the five different fitness functions for the
differential evolution tuning, it was necessary to execute differential
evolution five times, each time with another fitness function.  This is not
required for grid search, because the selection of the best can be done after
all raw data has been computed.  For \DEconfig{1}{1}, running the tuning lasted
for \qty{\miodeONEPLUSONESUMdays}{\day}; \DEconfig{1}{0} consumed
\qty{\miodeONEPLUSZEROSUMdays}{\day}, \DEconfig{0}{1} consumed
\qty{\miodeZEROPLUSONESUMdays}{\day}, \DEconfig{10}{1} consumed
\qty{\miodeTENPLUSONESUMdays}{\day}, and \DEconfig{1}{10} consumed
\qty{\miodeONEPLUSTENSUMdays}{\day}.  Still, the time required for grid search
is almost thrice the time required to run \emph{all} differential evolution.

For the \emph{DynaMOSA} tuning, grid search ran for \qty{\dynamosaGSdays}{\day}.
Differential evolution consumed \qty{\dynamosadeONEPLUSONESUMdays}{\day} for
\DEconfig{1}{1}, \qty{\dynamosadeONEPLUSZEROSUMdays}{\day} for \DEconfig{1}{0},
\qty{\dynamosadeZEROPLUSONESUMdays}{\day} for \DEconfig{0}{1},
\qty{\dynamosadeTENPLUSONESUMdays}{\day} for \DEconfig{10}{1}, and
\qty{\dynamosadeONEPLUSTENSUMdays}{\day} for \DEconfig{1}{10}.  Due to the
shorter runtime of grid search and the higher runtime of differential evolution,
compared to MIO, executing all five variants of differential evolution takes
longer than grid search; however, an individual run of differential evolution
only consumes between \perc{23.1} and \perc{49.0} of the time required for
grid search.

\rqsummary{RQ2}{%
  Differential evolution consumes significantly less time than grid search, while, at least for DynaMOSA, yielding better results.
}

\emph{Discussion:} The experiments show—in line with previous
research—that hyperparameter tuning is very resource intensive.  The
full experiment required a total runtime of \pynguin of
\qty{26281}{\day}, \ie, almost \exnum{72}~years, which is only
achievable with highly parallel computation.  For a cost estimation,
if we had executed the experiments on AWS-EC2 cloud services, where a
comparable machine costs US-\$\,\exnum{0.0384} per hour, it would have
resulted in a total cost of US-\$\,\exnum{24220.57}.
Consequently, tuning might not pay off if the test-generation
algorithms only seldom run, but for frequently running algorithms,
even small improvements can make a large difference over time.  We
believe that this is individual to the concrete use case and cannot be
decided in a study like this.

\section{Related Work}\label{sec:related}

Many studies indicate the necessity of hyperparameter tuning in various fields
and also its effectiveness~\cite{FMS16,FM17,VQ21}.  Some authors use random
search to tune hyperparameters~\cite{BB12,VQG+20,VQ21}, while we decided to use
an exhaustive technique, grid search, and a non-exhaustive technique,
differential evolution.
Hyperparameter tuning of EAs is a well-studied problem.  Aspects, such
as the methods to control and set their parameters~\cite{EMS+07} or
various methods to tune the parameter of a standard genetic algorithm
for continuous function optimisation~\cite{MRN14} have been explored.
While these works target different aspects than ours, they also show
that tuning the algorithm is a necessary task because its performance
is impacted by many factors, such as parallel programming, genetic
encoding, goal selection, or termination characteristics~\cite{MS20}.
These factors even indicate the future work for test-generation shall
incorporate more factors.

Closest to our work are the studies by Arcuri and Fraser~\cite{AF11}
and Kotelyanskii and Kapfhammer~\cite{KK14}: the former apply
parameter tuning to \evosuite and their results indicate that the
default parameter values in the literature perform reasonably well for
test generation, however, tuning is acceptable for researchers and can
lead to improved performance~\cite{AF11,AF13}.  The latter use the
sequential parameter optimisation toolbox~(SPOT)~\cite{Bar10} to tune
\evosuite with similar results as Arcuri and Fraser~\cite{KK14}.
Previous studies on \pynguin~\cite{LKF20,LKF23} used default
parameters, but our study suggests that better parameter-value choices
exist.

Parameter control is complementary approach to improve an algorithm's
performance: while tuning is applied before running the algorithm,
parameter control adjusts the parameter values during the algorithm's
runtime~\cite{EHM99}. Parameter control has also been applied in
search-based software
testing~\cite{RRV10,PCW11,AG22}.%
Since tuning is extremely time-consuming, parameter control may be a
more practical solution~\cite{EHM99,PRW10}. An alternative to reduce
the costs of tuning lies in predicting the possible performance gain
before applying tuning~\cite{zamani2021pragmatic}.

\section{Conclusions}\label{sec:conclusions}

We applied hyperparameter tuning to the DynaMOSA and MIO
test-generation algorithms in the \pynguin framework.  Our results
show that tuning MIO did not result in an improvement in terms of
coverage, independently of the tuning method used.  However, for
tuning DynaMOSA, we were able to achieve significant improvements in
terms of coverage over the default hyperparameter settings in
\pynguin.  Differential evolution yielded the best results overall
while requiring significantly fewer computational resources, compared
to grid search.
Future work will include comparing these findings with other tuning
techniques and popular tuning frameworks such as SPOT~\cite{Bar10},
ParamILS~\cite{HHL+09}, or irace~\cite{lopez2016irace}.

\bibliographystyle{splncs04}
\bibliography{master}

\begin{thebibliography}{10}
\providecommand{\url}[1]{\texttt{#1}}
\providecommand{\urlprefix}{URL }
\providecommand{\doi}[1]{https://doi.org/#1}

\bibitem{AG22}
Almulla, H.K., Gay, G.: Learning how to search: generating effective test cases through adaptive fitness function selection. Empir.\ Softw.\ Eng.  \textbf{27}(2), ~38 (2022). \doi{10.1007-S10664-021-10048-8}

\bibitem{Arc18}
Arcuri, A.: Test suite generation with the many independent objective {(MIO)} algorithm. Inf.\ Softw.\ Technol.  \textbf{104},  195--206 (2018). \doi{10.1016/j.infsof.2018.05.003}

\bibitem{AB14}
Arcuri, A., Briand, L.C.: A hitchhiker's guide to statistical tests for assessing randomized algorithms in software engineering. Softw.\ Test.\ Verification Reliab.  \textbf{24}(3),  219--250 (2014). \doi{10.1002/stvr.1486}

\bibitem{AF11}
Arcuri, A., Fraser, G.: On parameter tuning in search based software engineering. In: Proc.\ SSBSE. LNCS, vol.~6956, pp. 33--47. Springer (2011). \doi{10.1007/978-3-642-23716-4_6}

\bibitem{AF13}
Arcuri, A., Fraser, G.: Parameter tuning or default values? an empirical investigation in search-based software engineering. Empir.\ Softw.\ Eng.  \textbf{18}(3),  594--623 (2013). \doi{10.1007/s10664-013-9249-9}

\bibitem{Bar10}
Bartz{-}Beielstein, T.: {SPOT:} an {R} package for automatic and interactive tuning of optimization algorithms by sequential parameter optimization. CoRR  \textbf{abs/1006.4645} (2010)

\bibitem{Bel57}
Bellman, R.E.: Dynamic Programming. Princeton University Press (1957)

\bibitem{Bel61}
Bellman, R.E.: Adaptive control processes: a guided tour. Princeton University Press (1961)

\bibitem{BB12}
Bergstra, J., Bengio, Y.: Random search for hyper-parameter optimization. J.\ Mach.\ Learn. Res.  \textbf{13},  281--305 (2012). \doi{10.5555/2503308.2188395}

\bibitem{CGA18}
Campos, J., Ge, Y., Albunian, N., Fraser, G., Eler, M., Arcuri, A.: An empirical evaluation of evolutionary algorithms for unit test suite generation. Inf.\ Softw.\ Technol.  \textbf{104},  207--235 (2018). \doi{10.1016/j.infsof.2018.08.010}

\bibitem{DCF+15}
Daka, E., Campos, J., Fraser, G., Dorn, J., Weimer, W.: Modeling readability to improve unit tests. In: Proc.\ ESEC/FSE. pp. 107--118. {ACM} (2015). \doi{10.1145/2786805.2786838}

\bibitem{DMS16}
Das, S., Mullick, S.S., Suganthan, P.N.: Recent advances in differential evolution – an updated survey. Swarm Evol.\ Comput.  \textbf{27},  1--30 (2016). \doi{10.1016/J.SWEVO.2016.01.004}

\bibitem{EHM99}
Eiben, A.E., Hinterding, R., Michalewicz, B.: Parameter control in evolutionary algorithms. {IEEE} Trans.\ Evol.\ Comput.  \textbf{3}(2),  124--141 (1999). \doi{10.1109/4235.771166}

\bibitem{EMS+07}
Eiben, A.E., Michalewicz, Z., Schoenauer, M., Smith, J.E.: Parameter control in evolutionary algorithms. In: Parameter Settings in Evolutionary Algorithms, Studies in Computational Intelligence, vol.~54, pp. 19--46. Springer (2007). \doi{10.1007/978-3-540-69432-8_2}

\bibitem{FA11}
Fraser, G., Arcuri, A.: Evosuite: Automatic test suite generation for object-oriented software. In: Proc.\ ESEC/FSE. pp. 416--419. {ACM} (2011). \doi{10.1145/2025113.2025179}

\bibitem{FA13}
Fraser, G., Arcuri, A.: Whole test suite generation. {IEEE} Trans.\ Software Eng.  \textbf{39}(2),  276--291 (2013). \doi{10.1109/TSE.2012.14}

\bibitem{FM17}
Fu, W., Menzies, T.: Easy over hard: A case study on deep learning. In: Proc.\ ESEC/FSE. pp. 49--60. {ACM} (2017). \doi{10.1145/3106237.3106256}

\bibitem{FMS16}
Fu, W., Menzies, T., Shen, X.: Tuning for software analytics: Is it really necessary? Inf.\ Softw.\ Technol.  \textbf{76},  135--146 (2016). \doi{10.1016/j.infsof.2016.04.017}

\bibitem{FNM16}
Fu, W., Nair, V., Menzies, T.: Why is differential evolution better than grid search for tuning defect predictors? CoRR  \textbf{abs/1609.02613} (2016)

\bibitem{HS12}
Herlihy, M., Shavit, N.: The Art of Multiprocessor Programming. Elsevier (2012)

\bibitem{HLY20}
Huang, C., an~Xin~Yao, Y.L.: A survey of automatic parameter tuning methods for metaheuristics. {IEEE} Trans.\ Evol.\ Comput.  \textbf{24}(2),  201--216 (2020). \doi{10.1109/TEVC.2019.2921598}

\bibitem{HHL+09}
Hutter, F., Hoos, H.H., Leyton{-}Brown, K., Stützle, T.: {ParamILS:} an automatic algorithm configuration framework. J.\ Artif.\ Intell.\ Res.  \textbf{36},  267--306 (2009). \doi{10.1613/JAIR.2861}

\bibitem{KK14}
Kotelyanskii, A., Kapfhammer, G.M.: Parameter tuning for search-based test-data generation revisited: Support for previous results. In: Proc.\ QSIC. pp. 79--84. {IEEE} (2014). \doi{10.1109/QSIC.2014.43}

\bibitem{LPL+23}
Lemieux, C., Inala, J.P., Lahiri, S.K., Sen, S.: Codamosa: Escaping coverage plateaus in test generation with pre-trained large language models. In: Proc.\ ICSE. pp. 919--931. {IEEE} (2023). \doi{10.1109/ICSE48619.2023.00085}

\bibitem{lopez2016irace}
L{\'o}pez-Ib{\'a}{\~n}ez, M., Dubois-Lacoste, J., C{\'a}ceres, L.P., Birattari, M., St{\"u}tzle, T.: The irace package: Iterated racing for automatic algorithm configuration. Oper. Res. Perspect.  \textbf{3},  43--58 (2016). \doi{10.1016/j.orp.2016.09.002}

\bibitem{LF22}
Lukasczyk, S., Fraser, G.: Pynguin: Automated unit test generation for {Python}. In: Proc.\ ICSE Companion. pp. 168--172. {IEEE}/{ACM} (2022). \doi{10.1145/3510454.3516829}

\bibitem{LKF20}
Lukasczyk, S., Kroi\ss{}, F., Fraser, G.: Automated unit test generation for {Python}. In: Proc.\ SSBSE. LNCS, vol. 12420, pp. 9--24. Springer (2020). \doi{10.1007/978-3-030-59762-7_2}

\bibitem{LKF23}
Lukasczyk, S., Kroiß, F., Fraser, G.: An empirical study of automated unit test generation for {Python}. Empir.\ Softw.\ Eng.  \textbf{28}(2),  36:1--36:46 (2023). \doi{10.1007/s10664-022-10248-w}

\bibitem{MW47}
Mann, H.B., Whitney, D.R.: On a test of whether one of two random variables is stochastically larger than the other. The Annals of Mathematical Statistics  \textbf{18}(1),  50--60 (1947). \doi{10.1214/aoms/1177730491}

\bibitem{McM04}
McMinn, P.: Search-based software test data generation: A survey. Softw.\ Test.\ Verification Reliab.  \textbf{14}(2),  105--156 (2004). \doi{10.1002/stvr.294}

\bibitem{MRN14}
Montero, E., Riff, M., Neveu, B.: A beginner's guide to tuning methods. Appl.\ Soft.\ Comput.  \textbf{17},  39--51 (2014). \doi{10.1016/J.ASOC.2013.12.017}

\bibitem{MS20}
Mosayebi, M., Sodhi, M.: Tuning genetic algorithm parameters using design of experiments. In: Proc.\ GECCO. pp. 1937--1944. {ACM} (2020). \doi{10.1145/3377929.3398136}

\bibitem{NT10}
Neri, F., Tirronen, V.: Recent advances in differential evolution: a survey and experimental analysis. Artif.\ Intelli.\ Rev.  \textbf{33}(1-2),  61--106 (2010). \doi{10.1007/S10462-009-9137-2}

\bibitem{PKT15}
Panichella, A., Kifetew, F.M., Tonella, P.: Reformulating branch coverage as a many-objective optimization problem. In: Proc.\ ICST. pp. 1--10. {IEEE} Comp. Soc. (2015). \doi{10.1109/ICST.2015.7102604}

\bibitem{PKT18b}
Panichella, A., Kifetew, F.M., Tonella, P.: Automated test case generation as a many-objective optimisation problem with dynamic selection of the targets. {IEEE} Trans.\ Software Eng.  \textbf{44}(2),  122--158 (2018). \doi{10.1109/TSE.2017.2663435}

\bibitem{PCW11}
Poulding, S.M., Clark, J.A., Waeselynck, H.: A principled evaluation of the effect of directed mutation on search-based statistical testing. In: Proc.\ ICST Workshops. pp. 184--193. {IEEE} Comp. Soc. (2011). \doi{10.1109/ICSTW.2011.36}

\bibitem{PRW10}
Preuss, M., Rudolph, G., Wessing, S.: Tuning optimization algorithms for real-world problems by means of surrogate modeling. In: Proc.\ GECCO. pp. 401--408. {ACM} (2010). \doi{10.1145/1830483.1830558}

\bibitem{RRV10}
Ribeiro, J.C.B., Zenha{-}Rela, M., de~Vega, F.F.: Adaptive evolutionary testing: An adaptive approach to search-based test case generation for object-oriented software. In: Proc.\ NICSO. Studies in Computational Intelligence, vol.~284, pp. 185--197. Springer (2010). \doi{10.1007/978-3-642-12538-6_16}

\bibitem{SP97}
Storn, R., Price, K.V.: Differential evolution – {A} simple and efficient heuristic for global optimization over continuous spaces. J.\ Glob.\ Optim.  \textbf{11}(4),  341--359 (1997). \doi{10.1023/A:1008202821328}

\bibitem{VD00}
Vargha, A., Delaney, H.D.: A critique and improvement of the cl common language effect size statistics of {McGraw} and {Wong}. Journal of Educational and Behavioral Statistics  \textbf{25}(2),  101--132 (2000). \doi{10.3102/10769986025002101}

\bibitem{VQ21}
Villalobos{-}Arias, L., Quesada{-}López, C.: Comparative study of random search hyper-parameter tuning for software effort estimation. In: Proc.\ PROMISE. pp. 21--29. {ACM} (2021). \doi{10.1145/3475960.3475986}

\bibitem{VQG+20}
Villalobos{-}Arias, L., Quesada{-}López, C., Guevara{-}Coto, J., Martínez, A., Jenkins, M.: Evaluating hyper-parameter tuning using random search in support vector machines for software effort estimation. In: Proc.\ PROMISE. pp. 31--40. {ACM} (2020). \doi{10.1145/3416508.3417121}

\bibitem{zamani2021pragmatic}
Zamani, S., Hemmati, H.: A pragmatic approach for hyper-parameter tuning in search-based test case generation. Empir.\ Softw.\ Eng.  \textbf{26}(6), ~126 (2021). \doi{10.1007/s10664-021-10024-2}

\end{thebibliography}

\end{document}